\documentclass[journal=jacsat,manuscript=article, keywords]{achemso}

\usepackage{chemformula} 

\usepackage{amsmath}
\usepackage{amsfonts}
\usepackage{amssymb}
\usepackage{color}

\usepackage{changes}
\author{Rémy Pawlak}
\affiliation{Department of Physics, University of  Basel, Klingelbergstrasse 82, 4056 Basel, Switzerland}
\email{remy.pawlak@unibas.ch}
\author{Khalid N. Anindya}
\affiliation{Engineering Physics Department, Polytechnique Montr\'eal, Montr\'eal (Qu\'e), H3C 3A7, Canada}
\author{Toshiki Shimizu}
\affiliation{Department of Chemistry, The University of Tokyo, 7-3-1 Hongo, Bunkyo-ku, Tokyo 113-0033, Japan}
\author{Jung-Ching Liu}
\affiliation{Department of Physics, University of  Basel, Klingelbergstrasse 82, 4056 Basel, Switzerland}
\author{Takumi Sakamaki}
\affiliation{Department of Chemistry, The University of Tokyo, 7-3-1 Hongo, Bunkyo-ku, Tokyo 113-0033, Japan}
\author{Rui Shang}
\affiliation{Department of Chemistry, The University of Tokyo, 7-3-1 Hongo, Bunkyo-ku, Tokyo 113-0033, Japan}
\author{Alain Rochefort}
\affiliation{Engineering Physics Department, Polytechnique Montr\'eal, Montr\'eal (Qu\'e), H3C 3A7, Canada}
\author{Eiichi Nakamura}
\affiliation{Department of Chemistry, The University of Tokyo, 7-3-1 Hongo, Bunkyo-ku, Tokyo 113-0033, Japan}
\author{Ernst Meyer}
\affiliation{Department of Physics, University of  Basel, Klingelbergstrasse 82, 4056 Basel, Switzerland}
\email{ernst.meyer@unibas.ch}

\title{Atomically precise incorporation of BN doped rubicene into graphene nanoribbons}

\keywords{BN-doped graphene nanoribbons, on-surface reaction, scanning tunneling microscopy, atomic force microscopy, density functional theory}

\begin{document}

\begin{abstract}
  Substituting heteroatoms and non-benzenoid carbons into nanographene structure offers an unique opportunity for atomic engineering of electronic properties. Here we show the bottom-up synthesis of graphene nanoribbons (GNRs) with embedded fused BN-doped rubicene components on a Au(111) surface using on-surface chemistry. Structural and electronic properties of the BN-GNRs are characterized by scanning tunneling microscopy (STM) and atomic force microscopy (AFM) with CO-terminated tips supported by numerical calculations. The periodic incorporation of BN heteroatoms in the GNR leads to an increase of the electronic band gap as compared to its undoped counterpart. This opens avenues for the rational design of semiconducting GNRs with optoelectronic properties.
\end{abstract}

\section*{Keywords}
BN-doped graphene nanoribbon, on-surface reaction, scanning tunneling microscopy, atomic force microscopy, density functional theory

\section*{Introduction}
Tuning the band gap of atomically-precise graphene nanoribbons (GNRs) has garnered interest as a versatile route to promote exotic quantum or (opto)electronic properties.\cite{Chen2020} While graphene is a semi-metal, GNRs are generally semiconducting materials opening up interesting prospects as electronic components\cite{Wu2007} such as diodes or light emitting components.\cite{Chong2018,Ma2020} With the development of bottom-up on-surface reactions,\cite{Cai2010} the atomic structure of such extended polycyclic aromatic hydrocarbons (PAH) can indeed be precisely controlled allowing to tailor their electronic properties via quantum size effect, by incorporating topological defects,\cite{Rizzo2020,Pawlak2020} chemical dopants\cite{Kawai2015,Kawai2018,Pawlak2021,Sun2022} or non-benzenoid elements.\cite{Liu2019,Fan2019,Fan2021} This strategy requires a prior design of the molecular precursors in terms of shape and chemistry in order to guide surface-assisted chemical reactions towards the desired structures. 

Non-benzenoid polycycles in graphene, such as pentagon (cyclopentadiene) or hexagon (benzene) rings, are of particular interest to influence the physicochemical properties due to local strain and conjugation. Despite being composed of only carbon atoms, fullerenes have for instance unique photo-physical properties as a result of the electron-accepting ability of cyclopentadiene (CP) rings composing the C 60 structure. CP is able to aromatize by accepting electrons through aromatic 4n+2 stabilizations. Rubicene, a molecular fragment of C$_{70}$ composed of an indacene backbone, exhibits a similar electron affinity and
has already proven its suitability for organic semiconductor applications.\cite{Lee2014,Liu2016}

In addition, the incorporation of heteroatomic dopants within the GNR lattice remains a solid strategy to widen electronic band gaps. Among potential candidates, boron nitride
(BN) offers the advantage to be isoelectronic with carbon for similar structures, in particular hexagonal boron nitride (hBN) and graphene. In contrast to graphene, hBN possesses a wide band gap and interesting photophysical properties while having
a lattice mismatch of about 1.4$\%$ with the carbon lattice.\cite{Wang2017} In this context, incorporating a high-density of BN atomic dopants with atomic precision using on-surface chemistry might serve as unique opportunity to gain control over the magnetic and (opto)electronic properties of atomically-precise graphene boron-nitride hybrid nanomaterials.\cite{Fan2012,Anindya2022}
\begin{figure*}[t]
\centering
\includegraphics[width=15.50cm]{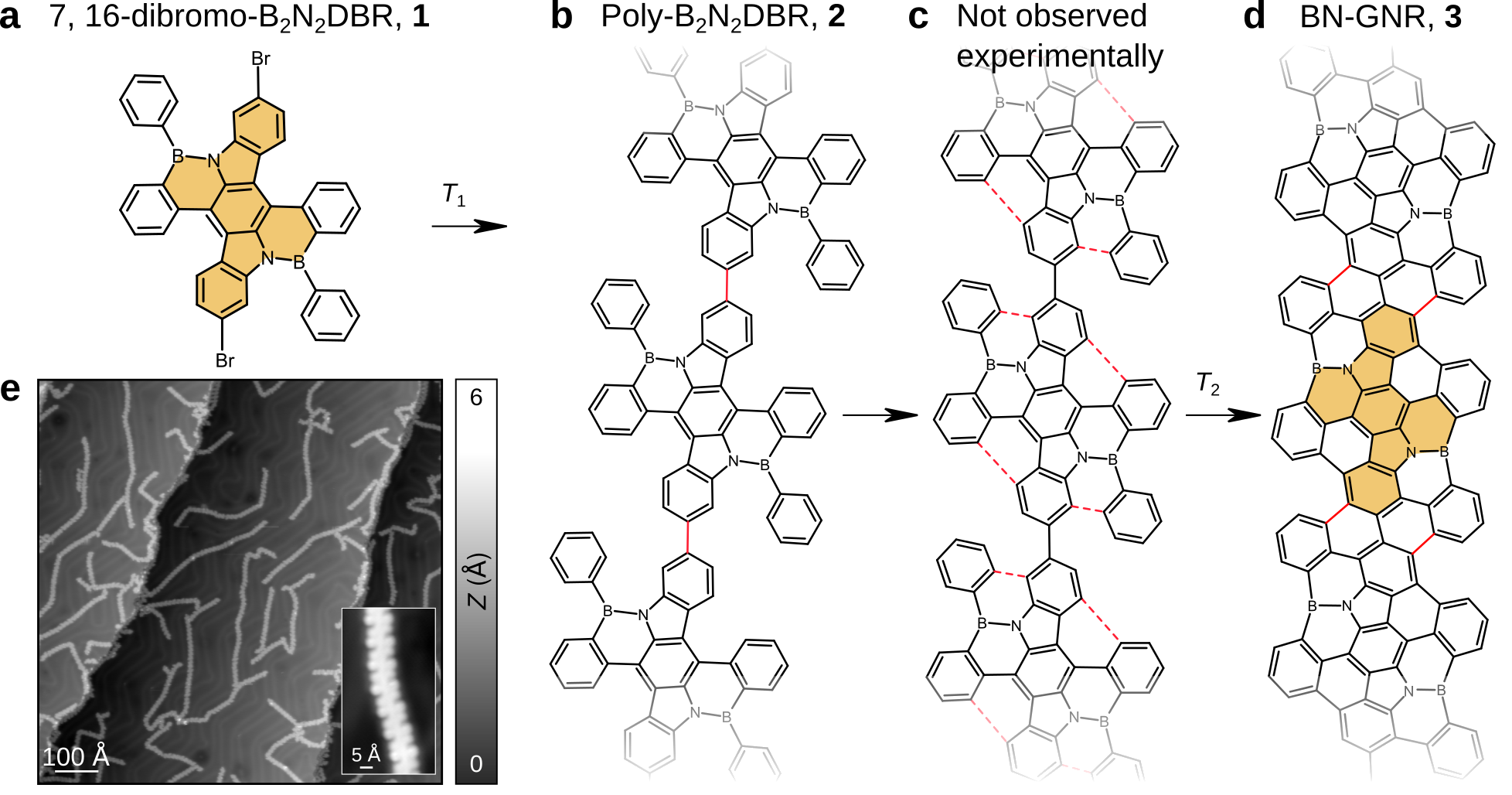}
\caption{Synthetic route to obtain BN-doped graphene nanoribbons (BN-GNR).
 {\bf a,} Structure of the B$_2$N$_2$-DBR precursor {\bf 1}. 
{\bf b,} Scheme of the reaction sequence obtained by annealing Au(111) to $T_1$ $\approx$ 160 $^\circ$C to initiate the Ullmann coupling reaction leading to poly-B$_2$N$_2$DBR polymeric chains {\bf 2}. 
{\bf c-d,} A subsequent substrate annealing to $T_2$ $\approx$ 250 $^\circ$C  induces cyclodehydrogenation reactions within monomers (dashed lines in {\bf c}) and between monomers (plain red lines in {\bf d}). This leads to the structure of BN-GNR {\bf 3} containing rubicene moieties (orange).
{\bf e,} STM overview of BN-GNRs adsorbed on Au(111). The inset shows a zoom STM image of a GNR.
} 
 \label{Fig1}
\vspace*{6mm}
\end{figure*}

Herein, we report on a surface-assisted synthesis of graphene nanoribbon (GNR) embedding fused BN-doped rubicene on Au(111). To fabricate long GNRs, halogen-
substituted 7,16-dibromo-5,14-dihydro-5a,14a-diaza-,14diboradibenzo[a, m]rubicene precursors (B$_2$N$_2$-DBR 1, Figure~\ref{Fig1}a) were synthesized in solution (Figure S1) following the procedure described in reference\cite{Sakamaki2021} and then sublimated on an Au(111) surface in ultra-high vacuum (UHV) conditions. Using a hierarchical Ullmann coupling reaction, we produced via substrate annealing polymeric chains of B$_2$N$_2$-DBR molecules ({\bf 2}, Figure\ref{Fig1}b) and GNRs ({\bf 3}, Figure~\ref{Fig1}d). The atomic structures and electronic properties were examined by STM, bond-resolved STM and AFM with CO-terminated tips at low temperature (4.8 K)\cite{Giessibl2003} combined with density functional theory (DFT) calculations. Importantly, successful on-surface synthesis of a BN-doped indacene-embedded GNR is demonstrated (Figure~\ref{Fig1}e). The rubicene-embedded graphene nanoribbons reported herein open avenues toward fabrication of novel non-benzenoid nanographenes.

\section*{Methods}
\paragraph{Sample Preparation.}
Au(111) single crystals were cleaned by several sputtering and annealing cycles in ultra high vacuum. 7,16-dibromo-5,14-dihydro-5a,14a-diaza-5,14-diboradibenzo[{\it a,m}]rubicene (B$_2$N$_2$-DBR) precursors were  sublimated in ultra-high vacuum (UHV) from a quartz crucible heated at 275$^\circ$C onto the sample kept at room temperature. Subsequent annealing of the substrate at appropriate temperatures initiate the formation of compounds {\bf 1} and {\bf 3} as discussed in the manuscript.
 
\paragraph{STM/AFM Experiments.}
The experiments were performed using a low-temperature STM/AFM microscope operated at T = 4.8 K in ultrahigh vacuum (p $\approx$ 1 $\times$ 10$^{-10}$ mbar). The force sensor is tuning fork based on a qPlus design~\cite{Giessibl2003} operated in the frequency-modulation mode (resonance frequency $f_0$ $\approx$ 25 kHz, spring constant $k$ $\approx$ 1800 N.m$^{-1}$, quality factor $Q$ $\approx$ 14000 and oscillation amplitude $A$ $\approx$ 0.5~\AA). The bias voltage was applied to the tip. STM images are taken in constant-current mode. AFM measurements were acquired in constant-height mode at V = 0 V. The tip mounted to the qPlus sensor consists in a 25 $\mu$m-thick PtIr wire, shortened and sharpened with a focused ion beam. A clean and sharp Au tip was then prepared at low temperature by repeated indentations into the surface. A functionalized CO tip was created by picking up a single CO molecule from the surface. Scanning tunneling spectroscopy (STS) data acquired at low temperature with the lock-in technique at 4.8~K ($A_{\rm mod}$ = 25 meV, $f$ = 531 Hz).

\paragraph{DFT Calculations.}
The DFT calculations were performed within the PBE functional limit\cite{Perdew1996} using the Quantum ESPRESSO package.\cite{Giannozzi2009,Giannozzi2017} Ion-electron interactions were considered through the Projector-augmented plane wave (PAW) method.\cite{Bloechl1994} The graphene nanoribbon (GNR) adsorbed on the Au(111) surface was modeled by an unit cell (see Figure S4) that contains a fixed three layers substrate with 96 Au atoms, and a GNR phase with 58 atoms. Both free-standing and adsorbed GNR models were optimized until energy and force convergence threshold reached 10$^{-6}$ Ry and 10$^{-4}$ Ry/\AA, respectively, with a $k$-point mesh of 1$\times$12$\times$1.\cite{Monkhorst1976} In addition, a $k$-point grid of 1$\times$26$\times$1 was used for band structure and properties calculations, with an energy cut-off value of 50 Ry. Bader charge analysis was carried out using the Bader code\cite{Tang2009,Sanville2007} to estimate the charge transfer between the nanoribbon and the Au(111) substrate. Calculated AFM images were obtained with the probe-particle model in which the Bader charges obtained from DFT calculations were considered in the simulations.\cite{Hapala2014}

\section*{Results and Discussion}
\paragraph{Synthesis of BN-doped graphene nanoribbon.}
To experimentally address each step of the reaction, we annealed step-wise the Au(111) substrate and systematically characterized the resulting structures by STM and AFM imaging at low temperature. Isolated precursors were first sublimated on the Au(111) sample located in the microscope. As the sample temperature never exceeds 15 K upon deposition, the diffusion of the molecules on the substrate is reduced leading to individual and unreacted B$_2$N$_2$DBR precursors. The STM image of Figure S2a  shows the intact precursor {\bf 1} with two bright dots corresponding to the peripheral phenyl rings attached to the B$_2$N$_2$DBR molecule. The corresponding AFM image (Figure S2b) obtained at constant-height with a CO-terminated tip (see Methods) similarly unveils these peripheral phenyls as bright protrusions. Note that the B$_2$N$_2$DBR backbone is not resolved since peripheral phenyls prevent stable AFM imaging conditions at closer tip-sample separations.
\begin{figure*}[t!]
\centering
\includegraphics[width=0.65\textwidth]{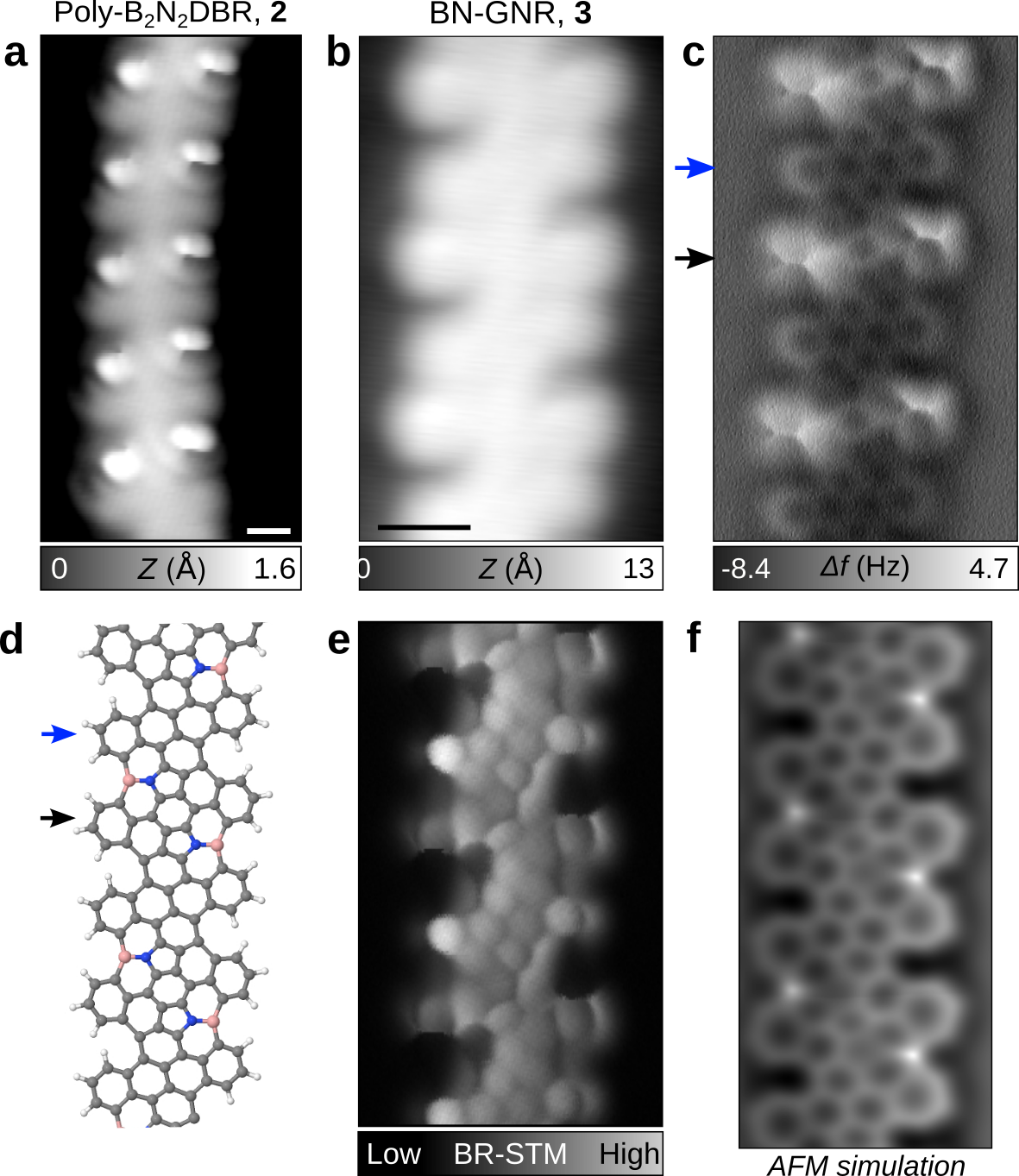}
\caption{Structural characterization of BN-GNR by AFM.
 {\bf a,} STM image of a poly-B$_2$N$_2$DBR chain {\bf 2}.
{\bf b,} STM image of a BN-GNR {\bf 3} ($I_{\rm t}$ = 1 pA, $V$ = 0.25 V), {\bf c,} Corresponding chemical structure by AFM and 
{\bf d,} Relaxed DFT structure of the BN-GNR. 
{\bf e,} Bond resolved-STM image with a CO-terminated tip. The black arrow points to the fused BN-doped rubicene and the blue arrow to the tetracene bridges. Scale bars are 5 \AA. The black arrow indicates the fused B$_2$N$_2$-DBR molecule incorporated into the GNR. 
{\bf f,} Simulated AFM image from the DFT coordinates.} 
 \label{Fig2}
 \vspace*{6mm}
\end{figure*}

Upon annealing to $T_1$ = 160 $^\circ$C, long poly-B$_2$N$_2$-DBR chains  are formed by Ullmann-coupling reaction (${\bf 2}$, Figure~\ref{Fig2}a). The detailed structure of this chain is revealed by a combination of high-resolution STM and constant-height AFM images (Figure~S2c and Figure S2b). Inter-molecular C-C bonds are created between each B$_2$N$_2$DBR monomer resulting in long polymeric chains. Similar to the intact precursor, bright protrusions are observed along the polymer edge that we attribute to the peripheral phenyls. We observe that the C-C bond formation occurs between monomers of identical orientations (red bonds in Figure~\ref{Fig1}b) leading to long polymeric chains while the reaction between opposite orientations seems less favored due to steric hindrance between the adjacent phenyls. A final annealing step up to $T_{2}$ = 250-300 $^\circ$C  further induces  C-C bond formation within the polymeric chain by cyclodehydrogenation reactions. The structure of Figure ~\ref{Fig1}c shows with dashed red lines the four ring closures within each monomer obtained during the reaction. Note however that such fused B$_2$N$_2$-rubicene chains have not been experimentally observed since the reaction is likely accompanied by the formation of additional C-C bonds between adjacent monomers (red lines in Figure~\ref{Fig1}d).

The final structure consists in a BN-doped GNR observed by STM (Figure~\ref{Fig2}b), where fused B$_2$N$_2$-DBR precursors (black arrow in AFM image of Figure~\ref{Fig2}c) are bridged by tetracene moieties (blue arrow). This results in the planarization of peripheral phenyls which is supported by DFT calculations of the BN-GNR structure in gas phase (Figure~\ref{Fig2}d) and adsorbed on Au(111) (Figure S4). This observation is also confirmed in the high-resolution AFM and bond-resolved STM images (Figure \ref{Fig2}c and Figure \ref{Fig2}e).  Both cases indicate a flat structure with the longitudinal direction being slightly rotated by 12-13$^\circ$ with respect to the Au(111) main directions. These structural details obtained from DFT along with the calculated Bader charge for the adsorbed model (see Methods section for details) were used to simulate AFM images with the probe-particle model\cite{Hapala2014} (Figure~\ref{Fig2}f). In comparison to the experimental AFM image (Figure~\ref{Fig2}c), we may notice the appearance of bright contrasts along the edges located at the BN sites while retaining the structure of the initial precursors on gold in the simulated AFM image. The slight discrepancies regarding the positions and magnitude of these bright contrasts observed between simulated and experimental AFM images may originate from a deviation in the Bader charges obtained from DFT that is implemented into the classical probe-particle model. Nevertheless, we think that the distinctive bright contrast in AFM images might originate from charge accumulations into the fused BN-rubicene moieties of the BN-GNR from the Au(111) substrate owing to the planarization of the peripheral phenyls. This negative charge accumulation leads to additional Pauli repulsion forces between tip and resulting in a brighter contrast as pointed out in a previous study.\cite{Liebig2020}

\begin{figure*}[t!]
\centering
\includegraphics[width=15.8cm]{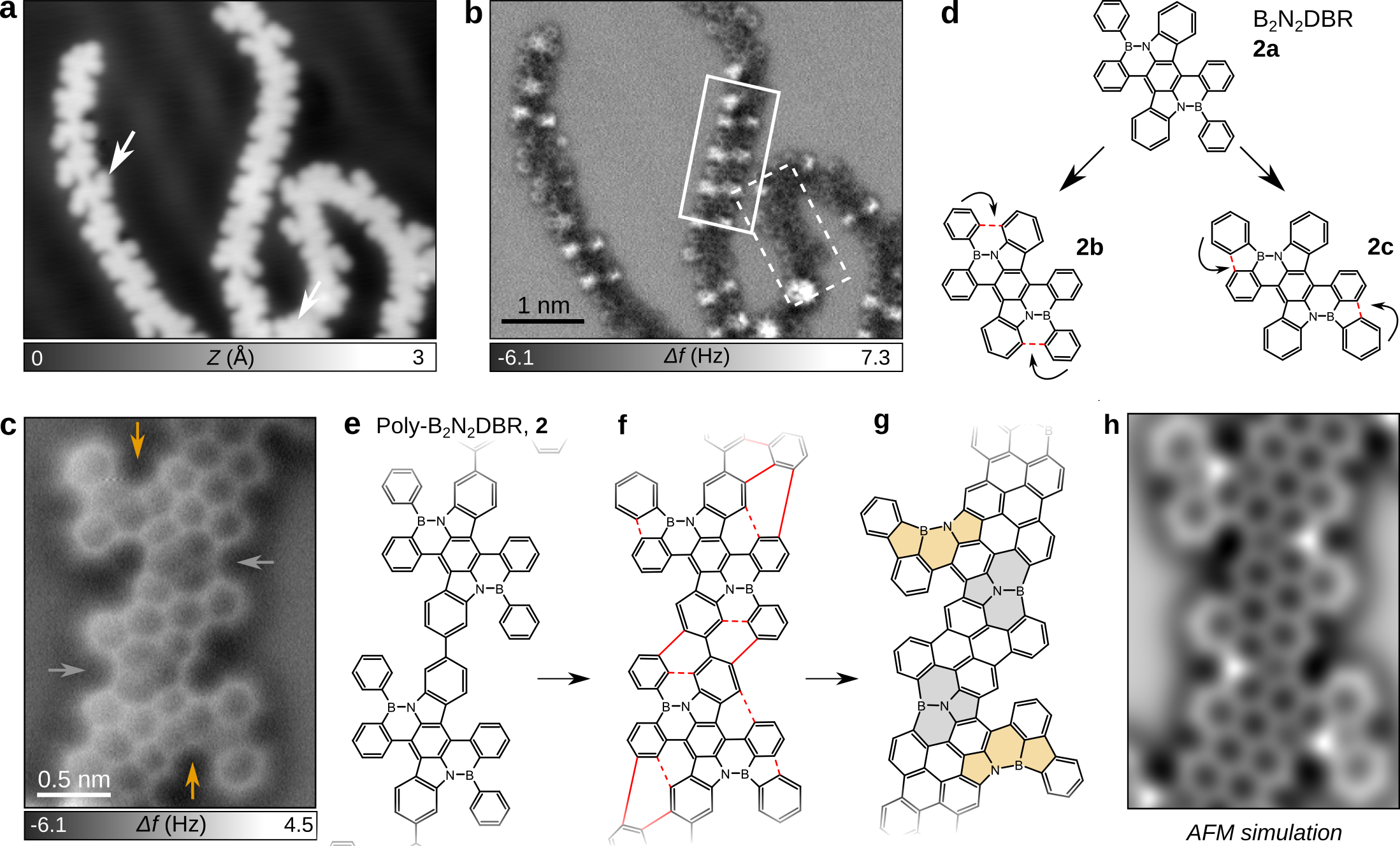}
\caption{Structural defects in BN-GNR. 
{\bf a,} Topographic STM image of BN-GNRs with structural defects exemplary marked by white arrows, ($I_{\rm t}$ = 1 pA, $V$ = 0.05 V). 
{\bf b,} Corresponding AFM image with a CO-terminated tip. The plain and dashed rectangles show BN-GNR and defected segments, respectively.   
{\bf c,} Close-up AFM image of the defected segment marked with a dashed square in {\bf b}. The gray and orange arrows refers to the BN atom sites. 
{\bf d,} Possible ring-forming reactions of a single B$_2$N$_2$DBR precursor {\bf 2a} leading to two distinct fused B$_2$N$_2$DBR molecules {\bf 2b} and {\bf 2c}.
{\bf e-g,} Proposed reaction from a poly-B$_2$N$_2$DBR chains leading to the defected segment shown in {\bf c}. Plain and dashed red lines in {\bf f} corresponds to  intermolecular and intramolecular C-C bond formation upon cyclodehydrogenation reaction, respectively. In {\bf g}, orange and gray area show the BN-atomic sites. {\bf h,} Simulated AFM image of the defected segment.
 \label{Fig2b}}
 \vspace*{6mm}
\end{figure*}
\paragraph{Structural defects of BN-GNRs.}
We next investigate structural defects of the BN-GNRs.  Experimentally, we found that the heating rate promote the formation of defects in the final BN-GNR structure.\cite{Jacobse2019} More precisely, after deposition of molecules on Au(111) kept at room temperature, a slow heating rate of about 30$^\circ$C/min from room temperature to $T_2$ leads to almost defect-free BN-GNR (Figure~\ref{Fig1}e, see also Figure S3).  In contrast,  higher heating rate of 100$^\circ$C/min up to the same maximum temperature $T_2$ tends to form more defected segments along the BN-GNR (shown in Figure~\ref{Fig2b}). (see Figure S3). Statistically, we found out that for slow heating rate ($\approx$ 30$^\circ$C/min) the number of defected segments along the BN-GNR is about 10-15 $\%$ (Figure S3) while, for higher heating rates ($\geq$ 100$^\circ$C/min), this value reaches up to 30-35 $\%$.

Figure~\ref{Fig2b}a and Figure~\ref{Fig2b}b show typical STM and AFM images of such structural peculiarity of GNRs (white arrows in Figure~\ref{Fig2b}a). The structure of pristine BN-GNR is still observed (plain rectangle in Figure~\ref{Fig2b}b) together with defected segments (dashed rectangle) which the chemical structure has been resolved by AFM in more details (Figure~\ref{Fig2b}c). As pointed out by gray and orange arrows, the BN locations appears darker than the carbon lattice with an increased contrast at the B atomic sites in excellent agreement with the AFM contrast of BN-doped GNRs.\cite{Kawai2018} Two distinct BN bonding configurations coexist in the structure, one being similar to the BN-GNR of Figure~\ref{Fig1}d (gray arrow) and one where BN atoms are at the ribbon edge (orange arrow). 

To better understand the synthetic pathway, we consider in Figure~\ref{Fig2b}d cyclodehydrogenation reactions of a single B$_2$N$_2$DBR molecule {\bf 2a} leading to two distinct fused B$_2$N$_2$DBR precursors {\bf 2b} and {\bf 2c}. Upon reaction, the peripheral phenyls can undergo ring-forming reactions resulting in either BN-doped hexagonal rings ({\bf 2b}) or B-doped pentagonal rings ({\bf 2c}). Starting from a poly-B$_2$N$_2$DBR chain (Figure~\ref{Fig2b}e), the formation of defected segments can be described by a combination of such ring formations within (dashed red bonds in Figure~\ref{Fig2b}f) and between monomers (plain red bonds) leading to the formation of the defected segment (Figure~\ref{Fig2b}g). Figure~\ref{Fig2b}h shows the simulated AFM image of such segment from the DFT coordinates of the relaxed structure. We emphasize that since the on-surface reaction for slow heating rates mostly leads to the BN-GNR structure, the {\bf 2b} compound is likely more stable than the {\bf 2c} one. To confirm this assumption, we have also calculated using DFT both structures and found out that molecule {\bf 2b} is energetically more stable than {\bf 2c} by around 1.4 eV. In analogy to reference,\cite{Bjoerk2022} we also think that the synthesis of {\bf 2c} might be favored by involving Au adatoms in the reaction during high heating rates. 

\begin{figure*}[t!]
\centering
\includegraphics[width=0.98\textwidth]{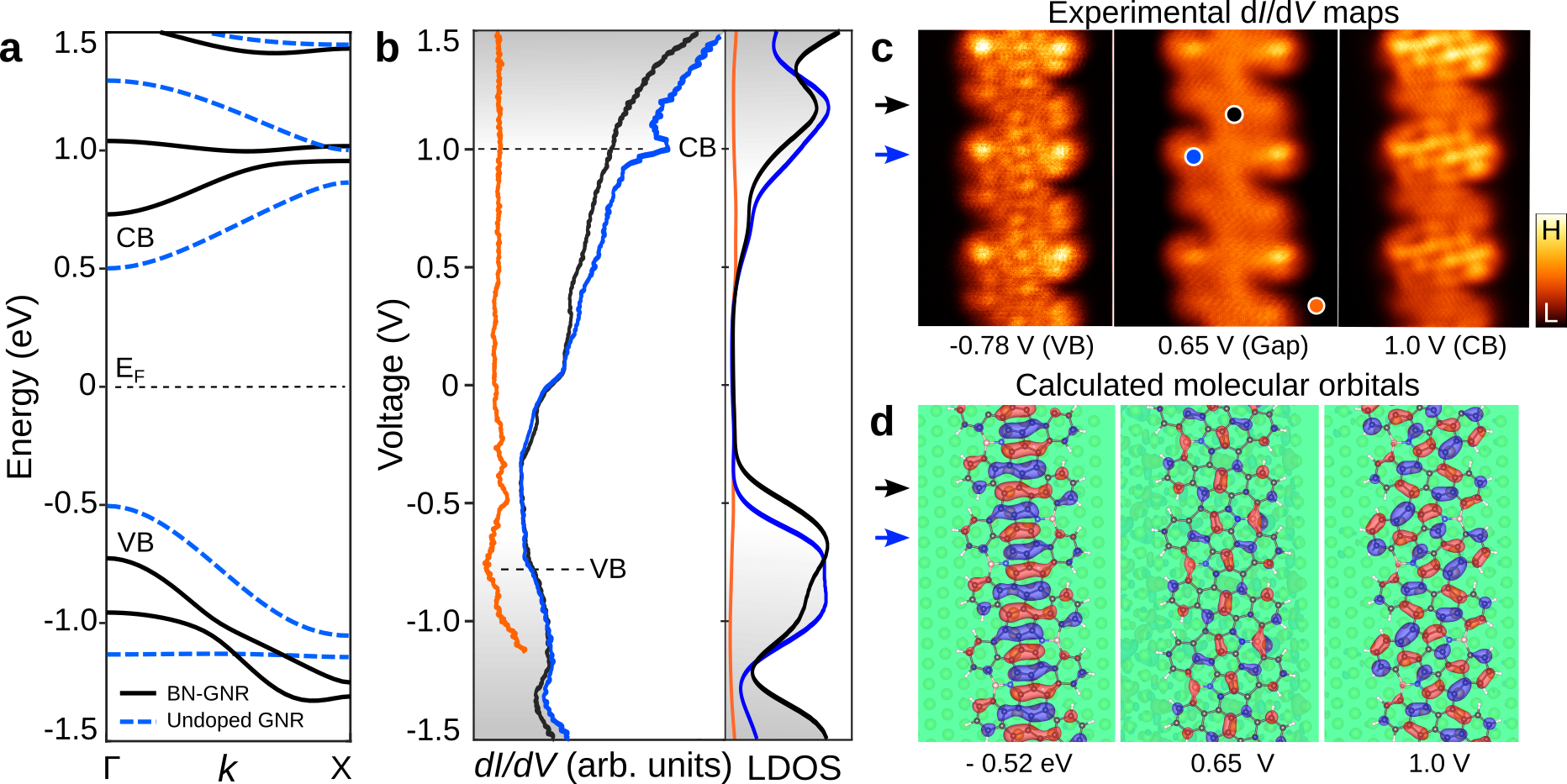}
\caption{Electronic structure of the BN-GNRs.
{\bf a,} Gas-phase band structure calculated by DFT of the BN-GNR (black) and its undoped counterpart (blue dotted line). 
{\bf b,} Differential conductance spectrum, d$I$/d$V$, acquired above the BN-GNR monomer (black and blue dots in {\bf c}). The band gap $E_{\rm gap}$ between valence (VB = -0.78 V) and conduction band CB onset (1.0 V) is about 1.7 eV.
{\bf c,} Experimental d$I$/d$V$ maps at different energies, respectively. 
{\bf d,} Calculated molecular orbitals of the BN-GNR adsorbed on Au(111).} 
 \label{Fig3}
 \vspace*{6mm}
\end{figure*}
\paragraph{Electronic structure of the BN-GNRs.}
We now discuss the electronic structure of BN-GNR investigated by scanning tunneling microscopy (STS) and DFT calculations. Figure~\ref{Fig3}a shows the band structure of the BN-GNR (black lines) and of its undoped counterpart (dashed blue lines), both in vacuum from their relaxed ground state geometry optimized by DFT. Compared to the undoped structure, the periodic incorporation of BN dopants leads to a net increase of the band gap of about 1 eV. The calculated band structure reveals the presence of well dispersed valence (VB) and conduction (CB) bands separated by a band gap of 1.55 eV (see Figure S5), which decreases to 1.46 eV upon adsorption on Au(111). Experimentally, differential conductance spectrum d$I$/d$V$ of Figure~\ref{Fig3}b were acquired above the fused B$_2$N$_2$DBR (blue), the tetracene bridge (black) and the gold surface (orange), which locations are also marked as colored circles in the spatially resolved d$I$/d$V$ maps (Figure~\ref{Fig3}c). The broad resonance peak centered at -1.0 V with an onset at -0.8 V is attributed to the VB edge, while the resonance peak only detected at +1 V above the fused rubicene (blue curve) is interpreted as the CB edge of the GNR.  Note that they are in relative agreement with the computed local density of states (LDOS) of the BN-GNR adsorbed on Au(111) (Figure~\ref{Fig3}b, Figure S5 and Figure S6) which indicates the VB and CB onsets at -0.52 eV and +1.07 eV, respectively.  

The d$I$/d$V$ map at the VB onset (-0.78 V) shows strong maxima over the fused B$_2$N$_2$DBR moiety (blue arrow) and a reduced charge density over the tetracene bridge (dark arrow). At the center of the GNR, a periodic modulation of the LDOS, resembling the HOMO of the rubicene precursors\cite{Sakamaki2021}, is 
delocalized along the polymer in remarkable agreement with the calculated molecular orbitals (-0.52 eV, Figure~\ref{Fig3}d). The d$I$/d$V$ map acquired at the CB edge (+1.0 V, Figure~\ref{Fig3}c) shows states mainly located over fused B$_2$N$_2$DBR moieties, again in line with the CB frontier orbitals (Figure~\ref{Fig3}d and Figure S6). Within the gap (exemplary shown at 0.65 V in Figure~\ref{Fig3}c), the d$I$/d$V$ maps show no site-dependent LDOS contrast that we attribute to the contribution of gold states through the molecule. 

We should last note that the band gap value obtained from STS measurements is typically reduced by electron screening from the metallic surface with respect to the intrinsic band gap of the gas phase polymer. Nevertheless, the remarkable agreement between experimental d$I$/d$V$ maps and DFT frontier orbitals validates the semiconducting character of the BN-GNR structure predicted by DFT. Thus, the BN-GNR has a large band gap with dispersive valence and conduction bands, which suggest future applications as semiconducting or opto-active elements.

\section*{Conclusion}
Our results introduce the on-surface synthesis on Au(111) of graphene nanoribbons from an optically-active precursor, the 7,16-dibromo-5,14-dihydro-5a,14a-diaza-5,14-diboradibenzo[$a,m$]rubicene (B$_2$N$_2$-DBR).\cite{Sakamaki2021} Upon reaction via Ullmann reaction, polymers of fused B$_2$N$_2$DBR molecules are formed that can be transformed into BN doped graphene nanoribbons via a thermally activated cyclodehydrogenation reaction. The BN-GNR structure identified by atomic force microscopy contains a periodic BN heteroatomic doping which open a large band of about 1.6 eV in the band structure as confirmed by combined STS measurements and DFT calculations. Future works will focus on the electronic decoupling of these BN-GNRs to investigate their optical properties.\cite{Lee2014} Indeed, BN-doping not only modulates the optical property of the carbon-based GNR, but also offers the possibility of controlling it by reversibly adding and removing an external electrophile as demonstrated for the B$_2$N$_2$-DBR precursor.\cite{Sakamaki2021} We thus envision our work will open new routes to incorporate non-benzenoid and heteroatomic dopants in conjugated nanographene with the prospects of steering novel electronic properties for optoelectronics and organic solar cells.

\section*{Supporting Information}
The Supporting Information is available free of charge at !!!

\noindent
Synthesis of 7,16-dibromo-5,14-dihydro-5a,14a-diaza-5,14-
diboradibenzo[$a$,$m$]rubicene precursors (Figure S1); 
additional topographic STM/AFM images (Figure S2);
topographic STM image of single B$_2$N$_2$-DBR chains (Figure S3);
structure of the optimized BN-GNRs adsorbed on Au(111) obtained by DFT (Figure S4);
band structure and molecular orbitals of free-standing BN-GNRs optimized by DFT (Figure S5);
projected density of states (PDOS) of the BN-GNR (Figure S6);
calculated local density of states (LDOS) at different positions along the BN-GNRs (Figure S7) (PDF)

\begin{acknowledgement}
Financial support from the Swiss National Science Foundation (SNF) and the Swiss Nanoscience Institute (SNI) is gratefully acknowledged. We also thank the European Research Council (ERC) under the European Union’s Horizon 2020 research and innovation programme (ULTRADISS Grant Agreement No. 834402) and supports as a part of NCCR SPIN, a National Centre of Competence (or Excellence) in Research, funded by the Swiss National Science Foundation (grant number 51NF40-180604).
A.R. and K.N.A. acknowledges the support from the Natural Sciences and Engineering Research Council of Canada (NSERC), and they are grateful to Calcul Québec and Compute Canada for providing computational resources.
\end{acknowledgement}

\bibliography{literature}%

\section*{Author contributions}
R.P., T.S., E.N. and E.M. conceived the experiments. T.S., R.S. and E.N designed and synthesized the monomer. R.P. performed the STM/AFM measurements. K.N.A. and A.R. performed the DFT calculations. R.P., K.N.A. and A.R. analyzed the data. R.P. wrote the manuscript. R.P., K.N.A., T.S., J.-C.L., T.S., R.S., A.R., E.N. and E.M. discussed on the results and revised the manuscript.

\section*{Competing interests}
The authors declare no competing financial interests.

\section{TOC Graphic}
\centering
\includegraphics[width = 8.5cm]{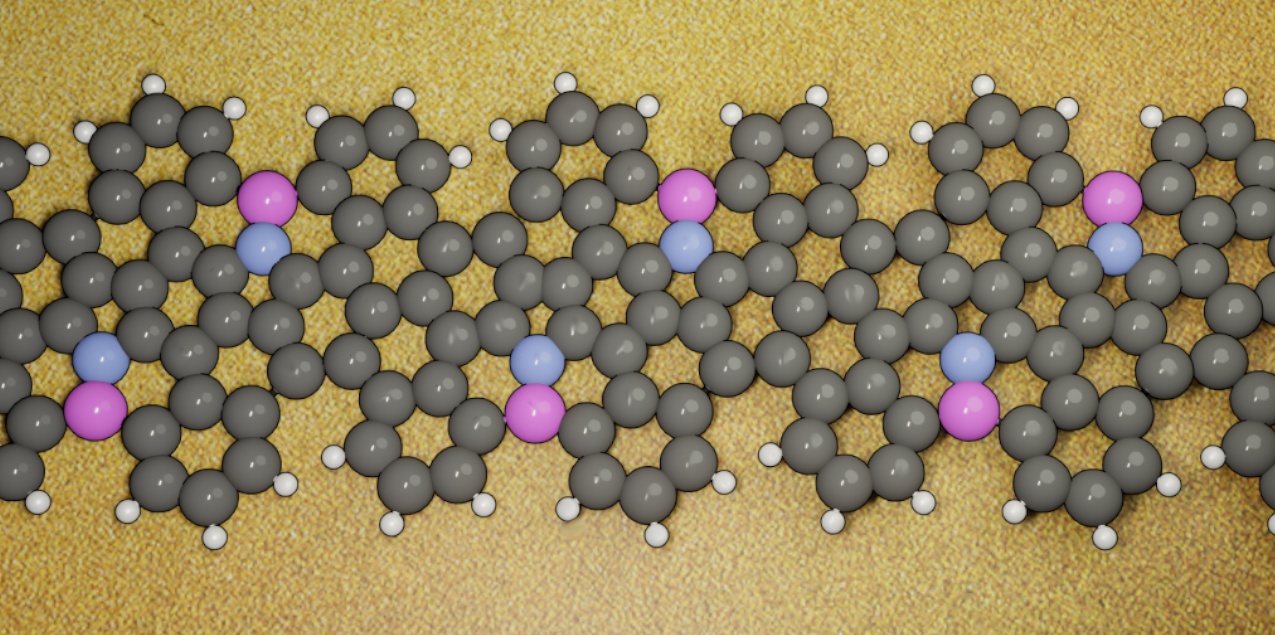}

\end{document}